\RequirePackage[l2tabu, orthodox]{nag}
\documentclass[conference, 10pt]{IEEEtran}

\usepackage[utf8]{inputenc}
\usepackage[T1]{fontenc}
\usepackage{newtx} 
\usepackage[american]{babel}
\usepackage{csquotes}
\usepackage{microtype}
\usepackage[dvipsnames]{xcolor}
\usepackage{graphicx}
\graphicspath{{./figures/}}
\usepackage{booktabs}
\usepackage[inline]{enumitem}
\usepackage{siunitx}
\usepackage{float}
\usepackage{algpseudocodex}
\usepackage[style=ieee]{biblatex}
\usepackage{orcidlink}
\usepackage{hyperref}
\usepackage[acronym]{glossaries} 
\glsdisablehyper 

\newacronym{bp}{BP}{Bundle Protocol}
\newacronym{cgr}{CGR}{Contact Graph Routing}
\newacronym{ctmc}{CTMC}{Continuous-Time Markov Chain}
\newacronym{dnf}{DNF}{Dependent Node Failures}
\newacronym{dtn}{DTN}{Delay-Tolerant Network}
\newacronym{lttg}{LTTG}{Latest Time to Goal}
\newacronym{mcts}{MCTS}{Monte Carlo Tree Search}
\newacronym[longplural={Markov Decision Processes}]{mdp}{MDP}{Markov Decision Process}
\newacronym{mtbf}{MTBF}{mean time between failures}
\newacronym{mttr}{MTTR}{mean time to repair}
\newacronym[longplural={Partially Observable Markov Decision Processes}]{pomdp}{POMDP}{Partially Observable Markov Decision Process}
\newacronym{rad}{RAD}{Risk-Aware Duplication}
\newacronym[longplural={Unobservable Markov Decision Processes}]{umdp}{UMDP}{Unobservable Markov Decision Process}
\newacronym{bruf}{BRUF}{Best Route Under Failures}
\newacronym{rucop}{RUCoP}{Routing under Uncertain Contact Plans}
\newacronym{lss}{LSS}{Lightweight Scheduler Sampling}

\hypersetup{
    pdftitle={POMDP-Based Routing for DTNs with Partial Knowledge and Dependent Failures},
    pdfauthor={Gregory F. Stock, Alexander Haberl, Juan A. Fraire, Holger Hermanns},
    pdfsubject={Space-Terrestrial Internetworking Workshop 2025}
}

\IEEEoverridecommandlockouts

\floatstyle{ruled}
\newfloat{algorithm}{tbp}{lop}
\floatname{algorithm}{Algorithm}

\newcommand\copyrighttext{%
    \footnotesize \textcopyright 2025 IEEE.
    Personal use of this material is permitted.
    Permission from IEEE must be obtained for all other uses, in any current or future media, including reprinting/republishing this material for advertising or promotional purposes, creating new collective works, for resale or redistribution to servers or lists, or reuse of any copyrighted component of this work in other works.
    DOI: \href{https://doi.org/10.1109/WiSEE57913.2025.11229850}{10.1109/WiSEE57913.2025.11229850}
}
\newcommand\copyrightnotice{%
    \begin{tikzpicture}[remember picture, overlay]
        \node[anchor=south, yshift=10pt] at (current page.south) {\fbox{\parbox{\dimexpr\textwidth-\fboxsep-\fboxrule\relax}{\copyrighttext}}};
    \end{tikzpicture}%
}

\begin{document}
\renewcommand{\figureautorefname}{Fig.} 
\renewcommand{\sectionautorefname}{Sec.} 
\renewcommand{\subsectionautorefname}{Sec.} 
\newcommand{\algorithmautorefname}{Algorithm} 

\title{POMDP-Based Routing for DTNs with Partial Knowledge and Dependent Failures
    \thanks{This project has received funding from the European Union's Horizon 2020 research and innovation programme under the Marie Skłodowska-Curie grant agreement \href{https://cordis.europa.eu/project/id/101008233}{No 101008233} -- \href{https://mission-project.eu}{MISSION} (\url{https://mission-project.eu}), and by DFG grant 389792660 as part of TRR~248 -- CPEC (\url{https://cpec.science}).}
}

\author{%
    \IEEEauthorblockN{%
        Gregory F. Stock\IEEEauthorrefmark{1} \orcidlink{0000-0001-5170-2019},
        Alexander Haberl\IEEEauthorrefmark{1} \orcidlink{0009-0000-6632-4276},
        Juan A. Fraire\IEEEauthorrefmark{1}\IEEEauthorrefmark{2}\IEEEauthorrefmark{3} \orcidlink{0000-0001-9816-6989}, and
        Holger Hermanns\IEEEauthorrefmark{1} \orcidlink{0000-0002-2766-9615}
    }
    \IEEEauthorblockA{\IEEEauthorrefmark{1}Saarland University -- Computer Science, Saarland Informatics Campus, Saarbrücken, Germany}
    \IEEEauthorblockA{\IEEEauthorrefmark{2}CONICET -- Universidad Nacional de Córdoba, Córdoba, Argentina}
    \IEEEauthorblockA{\IEEEauthorrefmark{3}Inria, INSA Lyon, CITI, UR3720, 69621 Villeurbanne, France}
}

\maketitle

\begin{abstract}
    Routing in \glspl{dtn} is inherently challenging due to sparse connectivity, long delays, and frequent disruptions.
    While \glspl{mdp} have been used to model uncertainty, they assume full state observability---an assumption that breaks down in partitioned \glspl{dtn}, where each node operates with inherently partial knowledge of the network state.
    In this work, we investigate the role of \glspl{pomdp} for \gls{dtn} routing under uncertainty.
    We introduce and evaluate a novel model: \gls{dnf}, which captures correlated node failures via repairable node states modeled as \glspl{ctmc}.
    We implement the model using JuliaPOMDP and integrate it with \gls{dtn} simulations via DtnSim.
    Our evaluation demonstrates that \gls{pomdp}-based routing yields improved delivery ratios and delay performance under uncertain conditions while maintaining scalability.
    These results highlight the potential of \glspl{pomdp} as a principled foundation for decision-making in future \gls{dtn} deployments.
\end{abstract}
\glsresetall 


\section{Introduction}

\copyrightnotice{}
The \gls{dtn} architecture was conceived to enable robust data transport over challenging environments such as near-Earth and deep-space networks~\cite{rfc4838}.
Communication in these networks is characterized by substantial delays due to vast distances, frequent disruptions caused by occlusions or duty-cycled equipment, together with the absence of persistent end-to-end paths.
The \gls{dtn} architecture addresses these challenges through a store-carry-and-forward approach, explicitly relaxing the assumption of continuous connectivity and immediate feedback from remote nodes.
The \gls{bp}~\cite{rfc9171} has emerged as the principal realization of \gls{dtn} concepts, with several operational implementations~\cite{DBLP:conf/infocom/BullDFKKSS24}.

Connectivity in \glspl{dtn} is represented through \textit{contacts}: finite time intervals during which two nodes can exchange data.
In space environments, these \textit{schedulable contacts} are often predictable with high accuracy, enabling so-called \textit{scheduled} \glspl{dtn}.
Contact opportunities can be determined in advance using orbital propagators and communication models, allowing operators to precompute a contact plan that encodes the future network topology.
This plan can then inform routing and forwarding decisions, whether centralized on the ground or distributed onboard spacecraft, forming the foundation for state-of-the-art algorithms such as \gls{cgr}~\cite{DBLP:journals/jnca/FraireJB21}.

However, the assumption of perfect knowledge of the contact plan is often unrealistic in practice.
Real-world scenarios introduce uncertainties such as interference, node failures, and environmental dynamics that can disrupt predicted contacts.
Prior works on \textit{probabilistic} \glspl{dtn}~\cite{rfc6693} fall short in this regard, as they typically do not account for scenarios with incomplete or inaccurate contact plans.
Extending scheduled routing in \gls{cgr} to properly model and react to such uncertainties through data replication has also proven challenging~\cite{DBLP:conf/wisee/BurleighCMR16}.

Recent work leveraging \glspl{mdp} has shown that this framework can enable effective multi-copy forwarding that improve delivery probability under uncertain contact conditions~\cite{DBLP:conf/wisee/RavertaDMFFD18,DBLP:journals/adhoc/RavertaFMDFD21}.
However, these models assume that the agent has perfect knowledge of the full network state, which is not the case in practical \glspl{dtn} with their partitioned and disconnected topologies.
Routing decisions in uncertain \glspl{dtn} must inherently operate under conditions of partial knowledge: each node acts based only on the knowledge of a pre-shared contact plan, local observations, and historical context, without access to a global, up-to-date view of the network state.
Despite efforts to approximate solutions locally using statistical~\cite{DBLP:conf/qest/DArgenioFHR22} and learning-based methods~\cite{DBLP:journals/tomacs/DArgenioFHR25}, the problem of formally addressing routing under uncertain contact plans with partial knowledge remains open.

This paper contributes a fundamental exploration of \glsdisp{pomdp}{Partially Observable MDPs (POMDPs)} as a foundation for routing in \glspl{dtn} where partial knowledge is intrinsic.
We introduce a novel model, \gls{dnf}, that captures node failures via repairable systems and transmission failures embedded within a \gls{pomdp} formulation.
This approach explicitly captures the uncertainty arising from both failure types and from each node's inherently partial, locally observed network view.
We implement the \gls{pomdp} using JuliaPOMDP~\cite{DBLP:journals/jmlr/EgorovSBWGK17} and integrate it with DtnSim~\cite{conf/smcit/FraireMRFV17} to enable simulation-based evaluation.
Extensive experiments show that \gls{pomdp}-based routing improves delivery ratio and energy efficiency, while remaining computationally feasible for onboard execution.

The remainder of this paper is structured as follows:
\autoref{sec:background} reviews the key concepts underlying this work.
\autoref{sec:case-study} presents our proposed \gls{dnf} model.
\autoref{sec:evaluation-and-results} evaluates \gls{dnf} through simulation experiments.
Finally, \autoref{sec:conclusion} summarizes our findings and outlines directions for future research.

\section{Background}\label{sec:background}

\subsection{Routing in \glsentrylongpl{dtn}}

\gls{cgr} is the preferred routing algorithm for scheduled \glspl{dtn} in space systems~\cite{DBLP:journals/jnca/FraireJB21}.
Its latest version was standardized as \gls{cgr}-SABR~\cite{standards/ccsds/SABR}.
\gls{cgr} operates under the assumption of a perfectly known contact plan and adapts Dijkstra's shortest path algorithm to compute time-optimal routes over a contact graph, a directed acyclic graph where vertices represent contacts, i.e.\@ unidirectional transmission opportunities from node~\(A\) to \(B\), and edges model storage delays between contacts.
\gls{cgr} selects the best route based on earliest delivery time and generally enforces a single-copy forwarding policy along this computed path---except for bundles flagged as critical~\cite{standards/ccsds/SABR}.

\gls{bruf}~\cite{DBLP:conf/wisee/RavertaDMFFD18} and \gls{rucop}~\cite{DBLP:journals/adhoc/RavertaFMDFD21} extend scheduled \gls{dtn} routing to handle contact plan uncertainties by framing routing as an \gls{mdp}.
\gls{bruf} computes an optimal single-copy policy that maximizes delivery probability under independent contact failures, serving as a theoretical upper bound for uncertain scenarios.
\gls{rucop} generalizes this to multi-copy forwarding, introducing practical variants: L-\gls{rucop}, a heuristic operating with local knowledge, and CGR-UCoP, which integrates uncertainty-aware decisions into standard \gls{cgr}.
A statistical variant~\cite{DBLP:conf/qest/DArgenioFHR22} leverages \gls{lss} and statistical model checking to estimate delivery probabilities from sampled local schedulers.
A learning-based variant~\cite{DBLP:journals/tomacs/DArgenioFHR25} applies Q-learning to incrementally train routing policies using simulation traces and local observations.
While all these methods improve delivery performance over \gls{cgr} when contact reliability is probabilistic and known a priori, they lack a formal treatment of routing under partial knowledge.

DtnSim~\cite{conf/smcit/FraireMRFV17} is a discrete-event simulator built on OMNeT++ specifically designed for evaluating routing protocols in \glspl{dtn}.
It integrates real flight-software implementations such as ION/\gls{cgr} and supports scheduled contact plans, enabling accelerated and reproducible simulations.
As a result, DtnSim has become the common platform for simulating bundle flows in many \gls{dtn} studies, including \gls{cgr}~\cite{DBLP:journals/jnca/FraireJB21}, \gls{bruf}~\cite{DBLP:conf/wisee/RavertaDMFFD18}, \gls{rucop}~\cite{DBLP:journals/adhoc/RavertaFMDFD21}, and their statistical~\cite{DBLP:conf/qest/DArgenioFHR22} and learning-based~\cite{DBLP:journals/tomacs/DArgenioFHR25} extensions.

\subsection{\glsentrylongpl{pomdp}}

The \gls{pomdp} model was introduced by \citeauthor{journals/maa/Astrom65}~\cite{journals/maa/Astrom65} as a generalization of \glspl{mdp} in which the agent is unable to observe the current state~\cite{journals/jmp/Littman09,DBLP:journals/ai/KaelblingLC98}.
Mathematically, a \gls{pomdp}~\(\mathcal{P}\) is defined as a 6-tuple \(\mathcal{P} = (S, A, Z, T, R, O)\), where \(S\) is a finite set of states, \(A\) is a finite set of actions, and \(Z\) is a finite set of observations.
\(T\colon S \times A \times S \to [0, 1]\) is the transition function, where \(T(s, a, s')\) indicates the probability to transition to state~\(s'\) when taking action~\(a\) in state~\(s\).
If action~\(a\) is \emph{enabled} in state~\(s\), it is a probability distribution over possible next states, i.e.\@ \(\sum_{s' \in S} T(s, a, s') = 1\) (otherwise \(\forall s' \in S. T(s, a, s') = 0\)).
For each state, at least one action must be enabled.
\(R\colon S \times A \times S \to \mathbb{R}\) is the reward function relating states and actions to real-value rewards, i.e.\@ \(R(s, a, s')\) indicates the expected reward for taking action~\(a\) from state~\(s\) and arriving in~\(s'\).
Note that, sometimes, the reward is independent of the resulting state (or the selected action).
Lastly, \(O\colon\! A \times S \times Z \to [0, 1]\) is the observation function: a probability distribution relating (resulting) states and (taken) actions to observations, i.e.\@ \(O(a, s', z)\) is the probability the agent will observe~\(z\) from taking action~\(a\) and landing in state~\(s'\).
We say that \(\mathcal{M} = (S, A, T, R)\) describes the \emph{underlying} \gls{mdp}.

Because the agent cannot directly observe the true state~\(s\) in a \gls{pomdp}, it instead maintains a \emph{belief state}~\(b\colon S \to [0, 1]\), which is a probability distribution over \(S\).
The initial belief is denoted as~\(b_0\), where \(b_0(s)\) is the probability that the system starts in~\(s\) for all~\(s\in S\).

Now, we show how the agent transitions from its current belief state~\(b\) to a new belief state~\(b'\) after executing an action~\(a\in A\) and receiving an observation~\(z\in Z\).
The (total) probability of observing~\(z\) is given by \(p_z = \sum_{s \in S} \left(b(s)\cdot\sum_{s' \in S} T(s, a, s') \cdot O(a, s', z)\right)\).
The new belief state takes into account the likelihood of ending up in state~\(s'\) from all states in~\(b\) and thereby observing~\(z\).
We can compute the new (normalized) belief state as \(b'(s') = \frac{1}{p_z}\sum_{s \in S} b(s) \cdot T(s, a, s') \cdot O(a, s', z)\).
Further, the expected reward of taking action~\(a\) is the average of possible rewards weighted by the probability of each state \(r_a = \sum_{s \in S} \sum_{s' \in S} b(s) \cdot R(s, a, s')\).

\subsubsection{Solving \glsentryshortpl{pomdp}}

Solving a \gls{pomdp} entails finding an optimal policy that maximizes the expected cumulative reward over time, based on belief states.
While \gls{mdp} policies map the history of states to actions, a \gls{pomdp} policy can only map the history of observations (or belief states) to actions.

Solving \glspl{pomdp} exactly (e.g.\@ using value iteration~\cite{DBLP:journals/ai/KaelblingLC98} or policy iteration) is computationally challenging due to the continuous and high-dimensional nature of the belief space.
However, approximate methods provide more scalable alternatives by trading off optimality for tractability, e.g.\@ sampling-based approaches such as Monte-Carlo planning~\cite{DBLP:conf/nips/SilverV10}.

\subsubsection{JuliaPOMDP}

JuliaPOMDP is a flexible ecosystem for modeling and solving \glspl{pomdp} in the Julia programming language.
We implement \gls{dnf} using POMDPs.jl~\cite{DBLP:journals/jmlr/EgorovSBWGK17}, which provides a common interface for formally specifying \gls{pomdp} models and supports a variety of solvers and simulation tools.
Specifically, we use the BasicPOMCP solver, which implements the PO-UCT online tree search algorithm~\cite{DBLP:conf/nips/SilverV10}.
PO-UCT is based on \gls{mcts} and approximates the optimal policy by (randomly) sampling trajectories.

\subsection{Repairable Systems}\label{sec:repairable-systems}

In the context of reliability engineering, a repairable system is a simple two-state model that alternates between an operational and a failed state~\cite{books/springer/Birolini17}.
The probabilities of transitioning between states are considered to be exponential distributions: the time until failure from the operational state is exponentially distributed with failure rate~\(\lambda\), while the time required for repair from the failed state is exponentially distributed with repair rate~\(\mu\).
The expected time spent in the operational state is called the \gls{mtbf}.
Similarly, the expected time spent in the failed state is called \gls{mttr}.
Since \(\lambda\) and \(\mu\) are constant, both values can immediately be derived as \(\mathit{\gls{mtbf}} = \frac{1}{\lambda}\) and \(\mathit{\gls{mttr}} = \frac{1}{\mu}\).

Formally, we can model a repairable system as a \gls{ctmc} with set of states~\(S = \{\top, \bot\}\), initial state~\(s_0 = \top\), and transition-rate matrix~\(Q = \left[\begin{smallmatrix} -\lambda & \lambda \\ \mu & -\mu \end{smallmatrix}\right]\).
From this, the transition matrix~\(P(t)\) can be computed, where~\(P_{ij}(t)\) is the probability that the system is in state~\(j\) at time~\(t\), given that it was in state~\(i\) at time~\(0\):
\[
    P(t) = \left[\begin{smallmatrix} P_{\top\top} & P_{\top\bot} \\ P_{\bot\top} & P_{\bot\bot} \end{smallmatrix}\right] = \mathrm{e}^{Qt} = \tfrac{1}{\lambda + \mu}\left(\left[\begin{smallmatrix} \mu & \lambda \\ \mu & \lambda \end{smallmatrix}\right] + \left[\begin{smallmatrix} \lambda & -\lambda \\ -\mu & \mu \end{smallmatrix}\right] \cdot \mathrm{e}^{-(\lambda+\mu)\cdot t}\right)
\]
For example, the probability of being in the operational state at time~\(t\) after initially having been in the operational state is given as \(P_{\top\top}(t) = \frac{\mu}{\lambda + \mu} + \frac{\lambda}{\lambda + \mu}\mathrm{e}^{-(\lambda+\mu)\cdot t}\).

Finally, we compute the stationary (or steady-state) distribution~\(\pi\)---the long-run probabilities of being in each state, independent of the initial state---by solving the equation \(\pi Q = \mathbf{0}\) under the constraint \(\pi_{\top} + \pi_{\bot} = 1\) and get \(\pi = \left[\begin{smallmatrix}\frac{\mu}{\lambda + \mu} & \frac{\lambda}{\lambda + \mu}\end{smallmatrix}\right]\).

\section{Case Study: \glsentrylong{dnf}}\label{sec:case-study}

This case study demonstrates how \glspl{pomdp} can be used to improve routing decisions in uncertain \glspl{dtn} by introducing a more realistic failure model.
Existing algorithms, such as \gls{rucop}~\cite{DBLP:journals/adhoc/RavertaFMDFD21}, are based on the contact failure model, which treats transmission failures as independent, i.e.\@ entire contacts from the original plan can fail with a given probability available a priori.
However, this model is unrealistic as all bundle transmissions during a contact either always fail or always succeed.
Instead, bundle transmissions to the same node should have dependent failure probabilities (i.e.\@ if the last transmission failed, the next one is more likely to fail for the same reason).

In the following, we introduce the \gls{dnf} model, originally developed in a Master's thesis by \citeauthor{thesis/uds/Haberl25}~\cite{thesis/uds/Haberl25}.
\gls{dnf} describes dependent node failures using repairable systems (see \autoref{sec:repairable-systems}) and, additionally, captures transmission failures---which are independent of node failures---caused by external factors such as noise.
Our scenario consists of a scheduled \gls{dtn} that uses destination-based routing (each intermediate node is solely responsible for determining the optimal next routing action) and a custody model~\cite{conf/spaceops/LeBihanFF25} (nodes need to know if a transmission was successful).
For simplicity, we assume that custody reports are always transmitted successfully.
Our \gls{pomdp} formulation computes such a single routing decision under the assumption of dependent node failures.
It is important to note that, although the current node simulates the bundle being routed through the entire network to the destination, it is unaware of the actual observations made by its neighbors.

More precisely, \gls{dnf} models the functional state of each node as a \gls{ctmc} (representing the repairable system).
Rather than being part of the state, the \gls{ctmc} is embedded in the \gls{pomdp}'s transition probability function, which mitigates the need to consider all functional states of nodes in each individual transition.
Instead, the transition probability function only has to consider the functional state of the \emph{receiving} node.
However, previous observations must be encoded in each state.
Observations represent previous transmission outcomes to a node and the time of the observation from the perspective of a local node.
The transition probability function then uses this information to predict the receiving node's functional state.

\subsection{\glsentryshort{dnf} State, Action, and Observation Spaces}

States of \gls{dnf} represent the network state (i.e.\@ temporal and node location of the bundle) and the current history of observations.
The network state is the observable part of the \gls{pomdp}, while the functional state of nodes is unobservable.
Importantly, uncertainty is not created by the node failure model itself, as it is embedded in the transition probability function.
Uncertainty is induced by the possibility of transmission failures, which make it uncertain whether the node or the transmission failed.
To define the \gls{pomdp} for \gls{dnf}, we first need to define the state, action, and observation spaces:

\subsubsection*{States}

A state in the \gls{pomdp} contains the network state of a bundle and an observation history.
It is represented as a tuple \(s = (n, t, \mathit{obs})\), where \(n\) is the node location of the bundle (identifier of a network node), \(t\) is the temporal location of the bundle (integer value; time is discrete starting from~\(0\)), and \(\mathit{obs}\) is the observation history mapping node identifiers to their last observed functional state and the observation time.
For example, \(s = (4, 3, \{2\colon \langle 1, \bot\rangle\})\) represents a state where the bundle is located at node~\(4\) at time~\(3\) and a single previous observation was made that node~\(2\) was observed as non-functional at time~\(1\).

\subsubsection*{Actions}

Actions correspond to schedulable contacts, including contacts in the distant future.
Note that, unlike in \gls{bruf}, there is no dedicated action to advance time.
Formally, \(a = c_{\mathit{id}}\) corresponds to a transmission using the contact with identifier~\(c_{\mathit{id}}\) from the contact plan.
Contacts are represented as tuples \(c_{\mathit{id}} = (\mathit{id}, \mathit{src}, \mathit{dst}, t_{\mathit{avail}}, t_{\mathit{prop}})\), containing the contact identifier~\(\mathit{id}\), source node~\(\mathit{src}\), destination (receiving) node~\(\mathit{dst}\), available time range (e.g.\@ \(t_{\mathit{avail}} = [50, 600]\)), and link propagation delay~\(t_{\mathit{prop}}\).

\subsubsection*{Observations}

We define two observations \(Z = \{z_{\top}, z_{\bot}\}\), representing the observed transmission outcome (success/failure).

\subsection{\glsentryshort{dnf} Failure Model}

The failure model provides the capability of predicting the success probability of a transmission under the assumption of node and transmission failures.
Here, we use the \gls{ctmc} representation of a repairable system as introduced in \autoref{sec:repairable-systems} and, for simplicity, assume that all nodes have a uniform \gls{mtbf} and \gls{mttr} and that all transmissions have the same failure probability~\(p^{\mathit{tx}}_{\mathit{fail}}\).
Therefore, only a single instance of the failure model is required for the entire network---more complex settings with individual failure models are also easily possible.

The failure model is initialized with \(\mathit{\gls{mtbf}}\) (or~\(\lambda\)), \(\mathit{\gls{mttr}}\) (or~\(\mu\)), and the transmission failure probability~\(p^{\mathit{tx}}_{\mathit{fail}}\).
Its core task is to predict the functional state of a receiving node given the current state of the \gls{pomdp} and an action (i.e.\@ a scheduled contact).
The probability that the receiver is in the functional state at the bundle arrival time is given by the transition matrix~\(P(\delta t)\), where \(\delta t\) is the (relative) elapsed time between the last observation to the arrival time---taking into account the current time and the link delay.
For example, assume the current state is \(s = (1, 15, \{2\colon \langle 10, \top\rangle\})\) and we select a contact that transmits the bundle from node~\(1\) to \(2\) with a link delay of \(3\).
Then, the model needs to predict the functional state of node \(2\) at time~\(18\), i.e.\@ the elapsed time is \(\delta t = 15 - 10 + 3 = 8\).
With \(\mathit{\gls{mtbf}} = 20\) (or \(\lambda = 0.05\)) and \(\mathit{\gls{mttr}} = 10\) (or \(\mu = 0.1\)), we can predict that the probability of node~\(2\) being functional at the arrival time is \qty{76.7}{\percent}.

For the (overall) success probability of a transmission, we have to multiply this value with the complement of the transmission failure probability \(1 - p^{\mathit{tx}}_{\mathit{fail}}\).

If there is no previous observation (i.e.\@ this is the first transmission to a node), \gls{dnf} assumes that the previous observation occurred in the indefinite past and uses the steady-state distribution~\(\pi\) instead.
This appears to be the best approach as \(\pi\) describes the long-run probabilities of being in each state, independent of the initial state (and time).

\subsubsection*{Optimization}

Evaluating the transition matrix~\(P(t)\) is one of the most computationally expensive processes in \gls{dnf}.
However, this process can be optimized without significantly compromising precision by exploiting that \(P(t)\) converges to~\(\pi\) for observations far in the past.
We therefore propose to use~\(\pi\) after a cut-off time, rather than computing the exact~\(P(t)\).
Given a user-defined precision~\(\varepsilon\), we can calculate the cut-off time~\(t_{\mathit{co}}\), after which the difference between~\(\pi\) and \(P(t)\) is less than~\(\varepsilon\).
For this, we solve \(P_{\top\top}(t_{\mathit{co}}) - \pi_{\top} < \varepsilon\) for \(t_{\mathit{co}}\), resulting in \(t_{\mathit{co}} > -\frac{1}{\lambda+\mu} \ln\left(\varepsilon \cdot \frac{\lambda + \mu}{\lambda}\right)\).

\subsection{\glsentryshort{dnf} Transition, Reward, and Observation Functions}

We now describe the remaining \gls{pomdp} components of \gls{dnf}.
Before we define \(T\), \(R\), and \(O\), we first define the initial and terminal states, and sketch the available action function required for the implementation.

\subsubsection{Initial Belief}

The initial belief \(b_0 = \langle s_0 \Rightarrow \qty{100}{\percent} \rangle\) contains a single state~\(s_0\).
This state contains the temporal and node location of the bundle that is currently routed.
If the current node has an observation history from previous transmission outcomes, it is added to the initial state.

\subsubsection{Terminal States and Actions}

Formally, \glspl{pomdp} do not have terminal states.
However, it is important to indicate to the solver when further progression is no longer meaningful, i.e.\@ when the bundle has reached its destination or becomes irreversibly stuck.
Since the value (and therefore further reward) of terminal states is always zero, we achieve this by introducing a special trap state along with a corresponding terminal action.

\subsubsection{Available Actions}

The available action function determines which contacts can be scheduled based on the given state~\(s\).
A contact is defined as \emph{schedulable} if its source node corresponds to the bundle location of \(s\) and if its available time range extends beyond the time of~\(s\).
Therefore, besides currently available contacts, contacts available in the future can also be scheduled.
Since the location and time of the bundle are fully observable, we know that all states in a belief share the same bundle location and time.
This implies that all states in the belief can schedule the same contacts.

The available action function is defined as follows:
First, given a belief~\(b\), select any state~\(s\) from \(b\).
If \(s\) is the terminal state, return the terminal action (every state must have at least one enabled action).
If the node location of~\(s\) is the destination node or the bundle cannot reach the destination node from~\(s\), return the terminal action.
The latter is determined using the \acrshort{lttg} heuristic to be introduced in \autoref{sec:lttg-heuristic}.
For all other cases, return all schedulable contacts from the contact plan.

The careful reader may have noticed that an action only specifies \emph{which} contact to use, not when.
In \gls{dnf}, contacts are used in a greedy way: we either forward the bundle immediately if the contact is already available or wait until the start of the contact.
Although there are edge cases where it would be better to wait until the end of a contact (e.g.\@ if \gls{mtbf} and \gls{mttr} are high and the receiver was last observed as non-functional), our approach is a trade-off to keep the state space tractable.

\subsubsection{Transitions}

The transition probability function, given a state~\(s\) and action~\(a\), determines a probability distribution over possible successor states.
Each action (except for the terminal action) has three successor states: a success state~\(s_s\), a \emph{node} failure state~\(s_f\), and a \emph{transmission} failure state~\(s_t\).

Assume the system is currently in \(s = (\mathit{src}, t, \mathit{obs})\) and an action with corresponding contact \(c_{\mathit{id}} = (\mathit{id}, \mathit{src}, \mathit{dst}, [t_s, t_e], t_{\mathit{prop}})\) is scheduled.
After a successful transmission, the bundle is at \(\mathit{dst}\).
Otherwise, it remains at \(\mathit{src}\) as required by the custody model.
To compute the new temporal location of the bundle, we consider the storage delay \(t_{\mathit{store}} = \max\{0, t_s - t\}\), i.e.\@ the time until~\(c_{\mathit{id}}\) becomes available, and the propagation delay~\(t_{\mathit{prop}}\).
This means that, in case of successful transmission, the bundle will be at the destination node at \(t_{\mathit{succ}} = t + t_{\mathit{store}} + t_{\mathit{prop}}\).
For failures, the custody timeout needs to be considered instead as it takes the same time for the acknowledgement message to return.
Only if there was no acknowledgement after \(t_{\mathit{fail}} = t_{\mathit{succ}} + t_{\mathit{prop}}+1\), we can safely deduce that the bundle was lost.

Finally, the observation history needs to be updated based on the transmission outcome.
For~\(s_s\) and \(s_f\), we update the entry of \(\mathit{dst}\) in \(\mathit{obs}\) to indicate that it was working/failed at the time of observation~\(t_{\mathit{succ}}\), respectively.
The transmission failure state~\(s_t\), however, is special as it has the node and temporal location of~\(s_f\) but receives the observation history of the success state~\(s_s\).
This is where partial observability is introduced into the model.
It is possible for a transmission to fail even though the receiving node is in a working state.
Formally, the possible successor states are \(s_s = \mathit{State(dst, t_{\mathit{succ}}, obs_{\top})}\), \(s_f = \mathit{State(src, t_{\mathit{fail}}, obs_{\bot})}\), and \(s_t = \mathit{State(src, t_{\mathit{fail}}, obs_{\top})}\), where \(\mathit{obs}_{\top} = \mathit{obs}[\mathit{dst}\mapsto \langle t_{\mathit{succ}}, \top\rangle]\) and \(\mathit{obs}_{\bot} = \mathit{obs}[\mathit{dst}\mapsto \langle t_{\mathit{succ}}, \bot\rangle]\).

The probability of the three states is determined as follows:
First, query the failure model for the probability of the destination node being in the failed state~\(p_{\bot}\).
This is the probability of transitioning to~\(s_f\).
Then, the probability of transitioning to~\(s_s\) or \(s_t\) is given by the product of the probability of the destination being operational~\(p_{\top}\) and the respective probability that a transmission failure occurred or not (i.e.\@ \(p_{\top} \cdot (1 - p^{\mathit{tx}}_{\mathit{fail}})\) for \(s_s\) and \(p_{\top}\cdot p^{\mathit{tx}}_{\mathit{fail}}\) for \(s_t\)).

\subsubsection{Observations}

The definition of the observation probability function is straightforward.
If the bundle was successfully transmitted (state \(s_s\)), a success observation~\(z_{\top}\) is made, and a failure observation~\(z_{\bot}\) otherwise (states \(s_f\) and \(s_t\)).

Note that the observation probability function is deterministic.
Yet, the observation influences how the current belief is updated.
If a success observation~\(z_{\top}\) was observed, we know that we must be in the success state~\(s_s\), while we cannot know in which of the failure states we are after seeing a failure observation~\(z_{\bot}\).

\subsubsection{Rewards}

This function rewards the agent for desired behavior.
Conceptually, we assign a positive reward when the destination is reached and a negative reward when the bundle is stuck.
More precisely, we define a negative reward \(\mathit{allStuck}\) if the bundle can no longer reach the destination (checked using the \acrshort{lttg} heuristic) and a time-adjusted reward \(\mathit{atGoal}\) if the bundle has reached the destination.
The latter decreases linearly with the arrival time to favor timely bundle delivery, i.e.\@ the agent receives the full reward if the bundle arrives at time~\(0\) and zero reward (still better than \(\mathit{allStuck}\)) if it arrives at the latest possible time.

\subsection{\glsentryfull{lttg} Heuristic}\label{sec:lttg-heuristic}

Determining if a bundle can still reach the destination is not trivial, yet it is valuable information required all across the model.
The \gls{lttg} heuristic can determine this property efficiently.
Notably, \gls{lttg} does not consider any link delays but only the available time range of contacts.
It is therefore a safe over-approximation for when it is certainly no longer possible to reach one node from another.

The \gls{lttg} heuristic is computed for all ordered pairs of nodes and stored as a matrix.
The computation of \gls{lttg} was inspired by \gls{rucop}~\cite{DBLP:journals/adhoc/RavertaFMDFD21} and begins its search from the destination node at the end of the time horizon.
The pseudocode is given in \autoref{alg:lttg}.
The backwards search allows the algorithm to ignore nodes that can never reach the destination and assign them a value of~\(-1\).

\begin{algorithm}[t]
    \caption{\glsentryfull{lttg} Heuristic.}\label{alg:lttg}
    \scriptsize%
    \begin{algorithmic}[1]
        \State \(\mathit{LTTG} \gets I_{\vert\mathit{nodes}\vert} -J_{\vert\mathit{nodes}\vert}\) \Comment{identity matrix~\(I\) minus all-ones matrix~\(J\)}
        \ForAll{\(\mathit{dst} \in \mathit{nodes}\)}
            \State \(\mathit{visited} \gets \{\mathit{dst}\}\) \Comment{set of visited nodes}
            \For{\(\mathit{time} = \mathit{endTime}, \ldots, \mathit{startTime}\)}
                \ForAll{\(\mathit{contact} \in \mathit{contactsAtTime(time)}\)}
                    \ForAll{\(\mathit{receivingNode} \in \mathit{visited}\)}
                        \If{\(\mathit{contact.dst} = \mathit{receivingNode} \wedge \mathit{contact.src} \notin \mathit{visited}\)}
                            \State \(\mathit{LTTG}[\mathit{dst}][\mathit{contact.src}] \gets \mathit{time}\)
                            \State \(\mathit{visited} \gets \mathit{visited} \cup \{\mathit{contact.src}\}\) \Comment{mark \(\mathit{contact.src}\) as visited}
                        \EndIf
                    \EndFor
                \EndFor
            \EndFor
        \EndFor
        \State \Return \(\mathit{LTTG}\) \Comment{\((\mathit{LTTG}[d][s] = t)\) \(\triangleq\) reaching~\(d\) from~\(s\) impossible after time~\(t\)\;} 
    \end{algorithmic}
\end{algorithm}

\subsection{\glsentrytext{dnf} Implementation and Application}

The \gls{dnf} model itself is implemented in JuliaPOMDP~\cite{DBLP:journals/jmlr/EgorovSBWGK17} and integrated in DtnSim~\cite{conf/smcit/FraireMRFV17} as a routing algorithm.
This means that, given a routing problem (current node and time of a bundle, destination, and observation history), DtnSim queries \gls{dnf} for each best next action.
\gls{dnf} then initializes a new \gls{pomdp} and initial belief each time, runs the solver to find the contact identifier over which the bundle should be routed, and returns this back to DtnSim.
Note that some initialization tasks (e.g.\@ parsing the contact plan, computing the \gls{lttg} heuristic, initializing the failure model) are performed just once and the results made available to all other (future) \gls{pomdp} instances.

\subsubsection*{Policy Determination}

As already stated, it is not possible to use an exact solver for \gls{dnf} as the state space explodes for realistic \glspl{dtn}.
Instead, we use BasicPOMCP, an online \gls{pomdp} solver based on \gls{mcts}.
As the solver depends on random sampling, the quality of the computed policy depends on the configured search parameters.
Notably, we currently use \num{10000} \gls{mcts} iterations, an exploration constant of \(c = 100\) (to favor exploration), and a maximum search depth of \(50\) (thus, \(50\) actions should suffice to reach the destination).

\subsubsection*{Estimator Function}

\gls{mcts}-based solvers require an estimator function which estimates the value function of a state.
The value function determines, given a state~\(s\) and policy~\(\pi\), the accumulated and discounted reward from~\(s\) and all future successor states, assuming that \(\pi\) is used.
To direct the \gls{mcts}, the estimator function should predict the value function as good as possible; however, an optimal prediction would require knowledge of the optimal policy, which is currently being approximated by \gls{mcts}.
While we plan to develop more advanced approach, we currently use the state's reward function as a heuristic for the estimator function.

\section{Evaluation and Results}\label{sec:evaluation-and-results}

We evaluated the performance of \gls{dnf} in minimal, hand-crafted scenarios (similar to unit tests) as well as in larger benchmarks with random contact plans using DtnSim.
Each minimal scenario tests a specific desired behavior, such as whether the model takes into account the arrival time of the bundle, the link delay, or previous observations when making routing decisions.

\begin{figure}
    \input{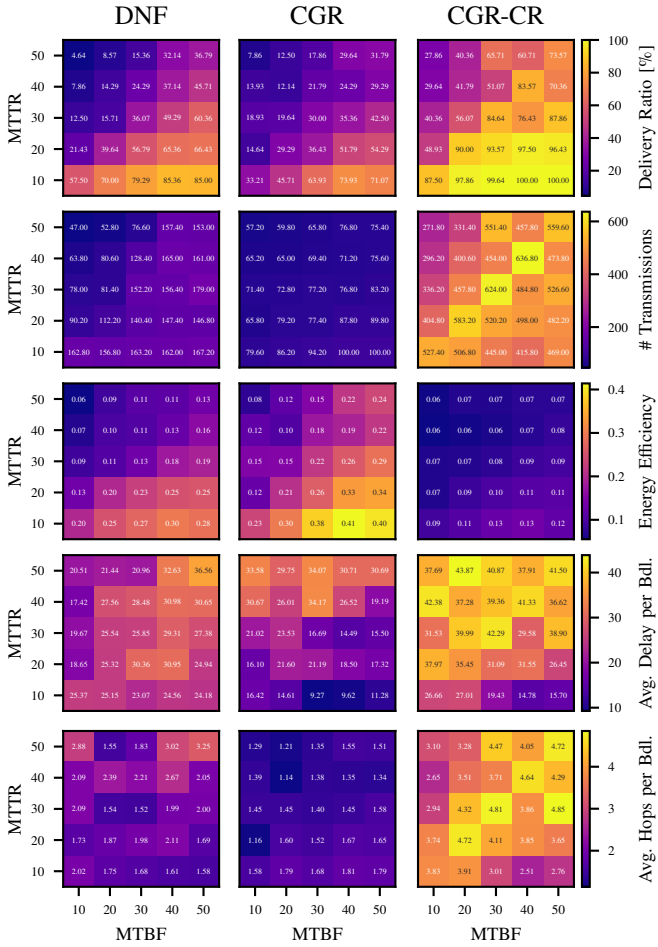}%
    \caption{Heatmaps comparing the metrics for \glsentrytext{dnf}, \glsentrytext{cgr}, and \glsentrytext{cgr}-CR under varying \glsentrytext{mtbf} \& \glsentrytext{mttr} and a transmission failure probability of \(5\%\).}\label{fig:eval-heatmaps}
\end{figure}

In the following, we take a detailed look at a general performance benchmark using a \qty{100}{\second} scenario with \(8\)~nodes and \(70\)~\emph{bidirectional} contacts, each having an available time range of \qty{10}{\second} and a link delay of \qty{2}{\second}.
All nodes were tasked with an all-to-all traffic pattern, i.e.\@ each node must transmit one bundle to all other nodes.
The collected performance metrics are the delivery ratio (percentage of generated bundles that are successfully delivered), total number of transmissions in the network, energy efficiency (number of successfully delivered bundles over number of transmissions), and end-to-end delay as well as hop count (mean delay \& hops of successfully delivered bundles; ignoring non-delivered bundles).
For each simulation, we selected the same \gls{mtbf} and \gls{mttr} for all nodes (between \qty{10}{\second} and \qty{50}{\second}), the same transmission failure probability for each transmission (\(0\%\), \(5\%\), or \(20\%\)), and repeated each scenario \(5\)~times with different random seeds to reduce noise.

As a baseline, \gls{cgr}~\cite{DBLP:journals/jnca/FraireJB21} was used in two flavors: (plain) \gls{cgr} without custody reports and \gls{cgr}-CR with custody reports~\cite{conf/spaceops/LeBihanFF25}.
All three algorithms are tested under the same dependent failure model (initialized with the same random seed).

The results of this benchmark series are visualized in \autoref{fig:eval-heatmaps}.
As expected, the delivery ratio is low if \gls{mtbf} is low and \gls{mttr} is high, implying a high probability of nodes being in a failed state.
Further, \gls{dnf} has an improved delivery ratio compared to \gls{cgr} but was still outperformed by \gls{cgr}-CR.
This shows that \gls{dnf} does not maximize delivery ratio but takes a compromise between energy efficiency, delivery delay, and hop count.
For example, \gls{dnf} is far more energy efficient than \gls{cgr}-CR because the latter aggressively tries to send the bundle over the lowest-delay
route (even if previous transmissions failed), while \gls{dnf} considers the functional state of the receiving nodes.
This aspect is even more apparent when looking at the number of transmissions.
Here, \gls{cgr} naturally has the lowest numbers since bundles are lost upon a single failed transmission.

When \gls{mttr} is high, \gls{dnf} experiences a high hop count, suggesting that \gls{dnf}, unlike \gls{cgr}, prefers rerouting to retransmission.
\gls{dnf} performs better than \gls{cgr}-CR in terms of delay and hops, showing that the observation history is beneficial to avoid spending time on routes with high failure probability.

In addition, we measured the running time and memory usage of each individual routing decision.
In this experiment series, it took a local node \(\qty{52}{\milli\second}\) (\(\sigma^2 = \qty{0.32}{\milli\second}\)) to compute one routing decision.
Note that the time is proportional to the number of \gls{mcts} iterations (here \num{10000}).
The average memory consumption was \qty{66}{\mega\byte}.

\section{Conclusion}\label{sec:conclusion}

This paper introduces a novel \gls{pomdp}-based routing model for uncertain \glspl{dtn} in the presence of dependent node failures.
The model shows how \glspl{pomdp} can reflect partial knowledge of local nodes.
Our evaluation results are encouraging: \gls{dnf} improves delivery ratios and energy efficiency, while remaining competitive in terms of execution time and delivery delay.

Nevertheless, several directions exist where further investigations could potentially improve the routing decisions.
We will extend this work by
\begin{enumerate*}[label=(\roman*)]
    \item exploring better estimator functions instead of using the reward function,
    \item sharing (actual) previous observations between (neighboring) nodes, and
    \item benchmarking the model on a wider variety of networks and comparing it with more advanced solutions like \gls{rucop}.
\end{enumerate*}

\renewcommand*{\bibfont}{\footnotesize}
\printbibliography

\end{document}